\documentclass[
 amsmath,amssymb,
 aps,
twocolumn]{revtex4-1}

\usepackage{xcolor}
\usepackage{graphicx}
\usepackage{dcolumn}
\usepackage{bm}
\usepackage{mathtools}

\newcommand{\wrlx}{\omega_\mathrm{rlx}}
\newcommand{\wsrv}{\omega_\mathrm{srv}}
\newcommand{\wflk}{\omega_\mathrm{flk}}

\newcommand{\wop}{\omega_\mathrm{opt}}
\newcommand{\wcut}{\omega_\mathrm{cut}}
\newcommand{\wlo}{\omega_\mathrm{LO}}

\newcommand{\SLO}{S_{\wlo}(\omega)}

\newcommand{\SO}{S_{\wop}(\omega)}
\newcommand{\SRIN}{S_\mathrm{RIN}(\omega)}

\newcommand{\Ssrv}{S_\mathrm{srv}(\omega)}

\begin{document}

\preprint{APS/123-QED}

\title{Limits on atomic qubit control from laser noise}

\author{Matthew L. Day}
\author{Pei Jiang Low}
\author{Brendan White}
\author{Rajibul Islam}
\author{Crystal Senko}
\affiliation{%
 Institute for Quantum Computing and Department of Physics and Astronomy, University of Waterloo, Waterloo, N2L 3R1, Canada\\
}%

\date{\today}

\begin{abstract}
{\color{black}
\noindent {Technical noise present in laser systems can limit their ability to perform high fidelity quantum control of atomic qubits. The ultimate fidelity floor for atomic qubits driven with laser radiation is due to spontaneous emission from excited energy levels. The goal is to suppress the technical noise from the laser source to below the spontaneous emission floor such that it is no longer a limiting factor. It has been shown that the spectral structure of control noise can have a large influence on achievable control fidelities, while prior studies of laser noise contributions have been restricted to noise magnitudes. Here, we study the unique spectral structure of laser noise and introduce a new metric that determines when a stabilised laser source has been optimised for quantum control of atomic qubits. We find requirements on stabilisation bandwidths that can be orders of magnitude higher than those required to simply narrow the linewidth of a laser. We introduce a new metric, the $\chi$-separation line, that provides a tool for the study and engineering of laser sources for quantum control of atomic qubits below the spontaneous emission floor.}}
\end{abstract}

\maketitle


\section{Introduction}
The laser has become an invaluable tool in the control of atomic systems due to electronic transitions in atoms typically having optical wavelengths. 
The application of the laser to the field of quantum information has been particularly effectual \cite{debnath2016demonstration,wright2019benchmarking,madjarov2020high,levine2019parallel}, and single atomic qubit control at the $10^{-4}$ error level has been demonstrated \cite{ballance2016high,pino2020demonstration}. 
The use of laser radiation for manipulating atomic energy levels will be fundamentally limited by spontaneous emission (SE), either due to the finite lifetime of qubits stored in optical transitions, or due to off-resonant scattering during two-photon Raman transitions. 
However, experimental demonstrations at the SE error floor have not been forthcoming due in part to technical noise sources dominating qubit errors. 
It is of interest to understand and reduce the technical error to the SE floor to enable low-overhead fault tolerant quantum computation. 

One dominant technical noise source is the local oscillator (LO) interacting with the qubit for quantum control. 
In this paper, we consider LOs derived from laser radiation. 
Previous studies have connected qubit fidelities to the total magnitude of laser noise \cite{wineland1998experimental,benhelm2008towards,schindler2013quantum,akerman2015universal,gaebler2016high,thom2018intensity}, and it has been demonstrated that the spectral structure of LO noise fields can have a critical influence in qubit fidelities through a study of phase noise in microwave sources \cite{ball2016role}. Here, we identify the conditions on the spectral structure of frequency and intensity noise of laser radiation to reduce these technical errors to, or below, the SE floor.

We find that, contrary to common belief \cite{akerman2015universal,bruzewicz2019trapped,yum2017optical}, narrowing the LO linewidth alone is not sufficient for high fidelity qubit control. 
The effective linewidth the qubit experiences is larger than the one given by a simple full-width-half-maximum (FWHM) measurement of the LO linewidth. This is because high frequency sideband noise on the LO carrier can have a sizeable effect on qubit fidelities, and stabilisation techniques are limited in their control bandwidth \cite{fluhmann2015spectral,stoehr2006diode,kirchmair2006frequency,chang2019stabilizing}. 

We find that laser frequency noise is a primary consideration, as we show that the qubit infidelity from shot-noise limited laser intensity noise is always below the SE floor, across all commonly-used atomic species and qubit types. 
In practice, laser sources are rarely shot noise limited and we outline requirements on intensity noise  stabilisation bandwidths to suppress these errors below the SE floor.

The results presented in this manuscript provide a roadmap to suppress qubit errors from technical laser noise to below the fundamental limit of atomic spontaneous emission.
The findings guide the choice of laser source and requirements for laser stabilisation. 
We find that for a stabilised laser source, there are three primary regimes. In the first regime, the stabilisation is insufficient to reduce qubit control errors. In the second regime, the stabilisation reduces control errors, but the unstabilised noise still dominates the error. In the third regime, the stabilisation is sufficient such that the errors are dominated by the stabilised noise amplitude. We develop a new metric, called the $\chi$-separation line, that determines when the third regime is satisfied, and that can be easily used to analyse realistic laser noise spectra. Critically, the $\chi$-separation line places stricter requirements on stabilisation loops than those for simply narrowing the laser linewidth. For a laser operating in the third regime we outline the requirements to operate below the SE floor for both laser frequency and intensity noise and find no fundamental obstacles to this goal.

\section{Results}

\subsection{Background} 

\begin{figure*}
    \centering
    \includegraphics[width=0.85\linewidth]{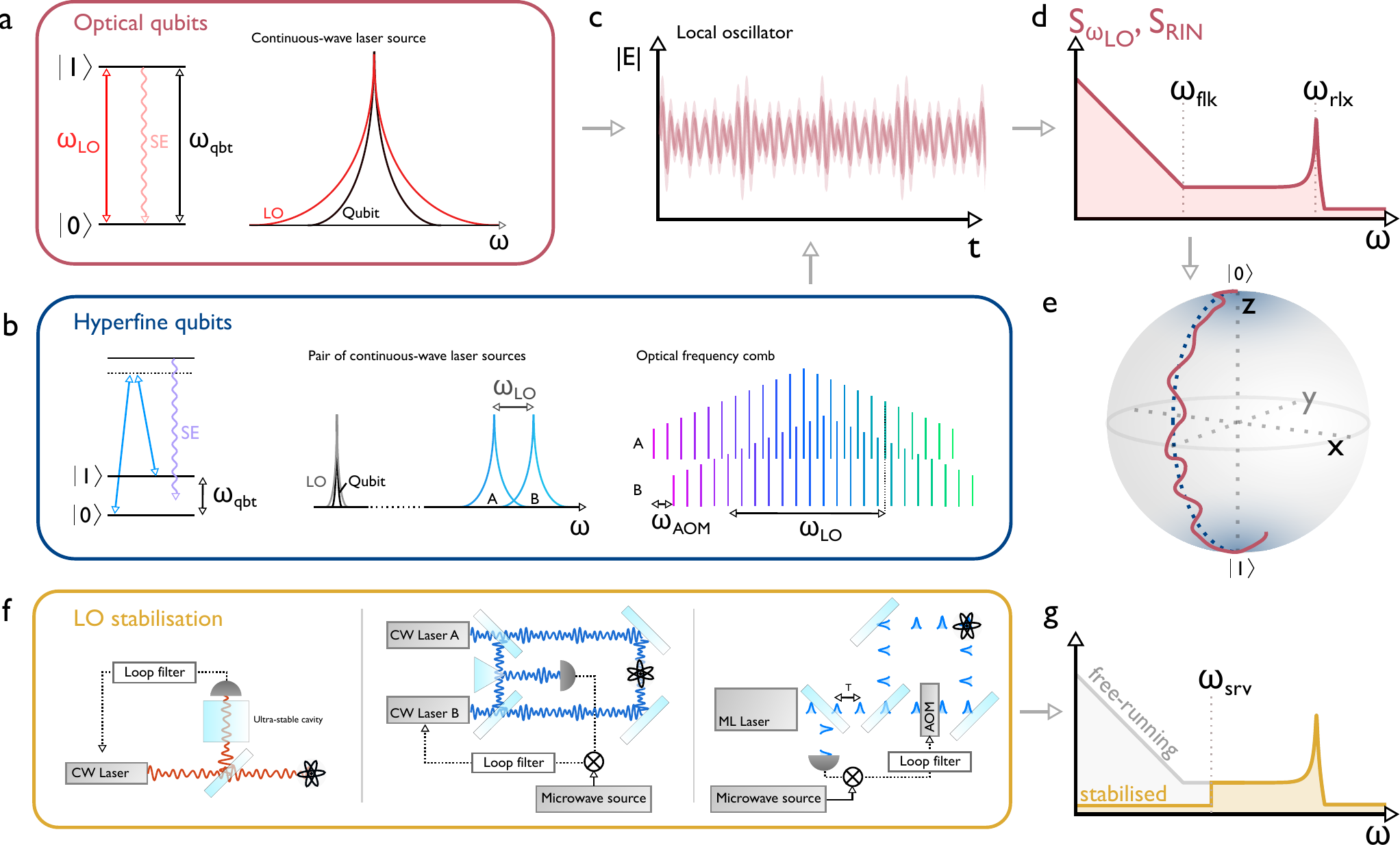}
    \caption{\textbf{Atomic qubit control using lasers as the local oscillator (LO)}.
    (a) For optical qubits, the laser light is resonant with the qubit transition. 
    (b) For hyperfine qubits, the LO is derived from the beatnote between two closely spaced optical frequencies.
    For continuous-wave (CW) lasers, this is performed by off-set phase locking two lasers by the qubit frequency. 
    For modelocked (ML) lasers, this is performed by interfering two frequency combs at the atom position, with one comb shifted in frequency such that the qubit is driven by the frequency difference between comb pairs. 
    (c) Noise away from the laser carrier frequency leads to a time-dependent variation in the LO control field, which can itself be expressed as a power spectral density, as shown in (d).
    (e) The time-dependent noise causes perturbations in qubit evolution on the Bloch sphere. (f) Example servo loops for stabilising laser frequency noise. (g) The servo loops act to reduce noise within the servo bandwidth, while free-running noise at higher frequencies is unaffected.}
    \label{fig:overview}
\end{figure*}

 Atomic qubits are typically encoded in energy levels separated by either optical or microwave transition frequencies. 
 We refer to these as optical and hyperfine qubits respectively.
 For optical transitions (Figure \ref{fig:overview}(a)), narrow-linewidth laser radiation can be directly used to coherently rotate the qubit state \cite{akerman2015universal}.
 For microwave transitions (Figure \ref{fig:overview}(b)), two phase-coherent optical fields offset by the qubit frequency can be used to control the qubit state through a two-photon Raman transition \cite{ballance2016high,mizrahi2014quantum}. 
One way to generate the beatnote for hyperfine qubit control is to phase-lock two continuous-wave (CW) lasers and interfere them at the atom position. 
A more recent approach is to use the output of a mode-locked (ML) laser. 
The pulse train of an ML laser forms a comb of optical frequencies, with each comb tooth separated by the repetition rate of the laser \cite{stowe2008direct}.
The mode-locking process ensures all comb teeth are phase coherent, and when the comb is split into two paths and interfered at the atom position, a series of beatnotes are generated. 
By placing a frequency shifter, such as an acousto-optical modulator (AOM), in one arm of the interferometer, beatnote harmonics can be finely tuned to the qubit frequency \cite{islam2014beat,mizrahi2014quantum}.

Noise in the laser sources used for optical and hyperfine qubit control (Figure \ref{fig:overview}(c-d)) leads to noisy evolution of the qubit on the Bloch sphere (Figure \ref{fig:overview}(e)).
Frequency noise causes unwanted rotations around the $Z$ axis, while intensity noise causes unwanted rotations around the $X$ and $Y$ axes. 
The accumulation of these perturbations leads to an imperfect overlap of the final quantum state with the target state. 
The overlap can be quantified by the fidelity, which can be calculated using filter function theory. For sufficiently small noise, the fidelity can be expressed as
\cite{ball2015walsh,kabytayev2014robustness,green2013arbitrary}
\begin{equation}
 \mathcal{F}^{(u)}(\tau)=\frac{1}{2}\left[1+\exp(-\chi^{(u)})\right]\approx 1-\frac{\chi^{(u)}}{2}~,   \label{eq:FF_fidelity_maintext}
\end{equation}
where the fidelity decay constant, $\chi^{(u)}$, is a spectral overlap of the laser noise power spectral density (PSD) with the filter function of the target operation, $u$, of duration $\tau$ (see Methods). 
In this paper we derive our results for general filter functions and present examples for a primitive $\pi$-pulse between the ground and excited state in an ideal two-level atom with a Rabi frequency $\Omega=\pi/\tau$. 
The fidelity of a $\pi$-pulse is chosen for ease of interpretation, and we note that fidelities of other common operations, such as a Ramsey sequence, are within an order of magnitude of the $\pi-$pulse for the same operation times \cite{ball2016role}.

The connection of the laser noise PSDs to the operation fidelity motivates the suppression of noise in the laser sidebands. 
As shown in Figure \ref{fig:overview}(f-g) for frequency noise, this is typically performed using active stabilisation (servo) loops of a finite bandwidth. 
We use Equation \ref{eq:FF_fidelity_maintext} to investigate the requirements on these stabilisation loops for maximising operation fidelities.

\subsection{Models of laser noise}

\begin{figure}
    \centering
    \includegraphics[width=1\linewidth]{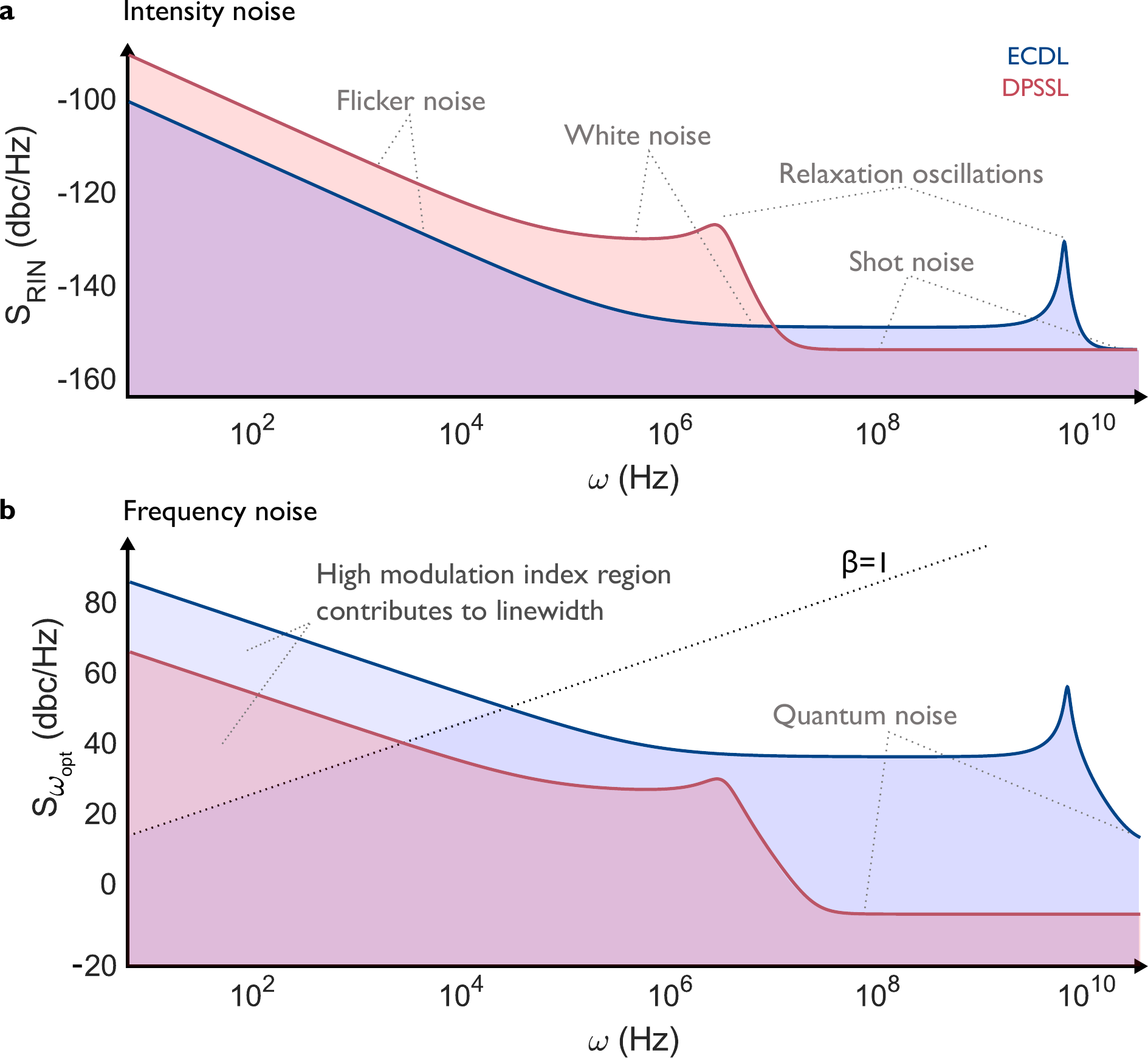}
    \caption{\textbf{Idealised free-running laser noise PSDs for external cavity diode lasers (ECDL) and diode-pump solid-state lasers (DPSSL).} (a) The RIN PSD, demonstrating the low frequency $1/\omega$ (flicker) noise, mid-frequency white noise, the relaxation oscillation peak, and shot noise. (b) The frequency noise PSD with the same general features as RIN with a fundamental limit of quantum noise. The noise above the $\beta$-separation line contributes to the linewidth of the laser, while the noise below the $\beta$-separation line contributes to the wings of the laser lineshape. }
    \label{fig:laser_noise_models}
\end{figure}

Noise processes in laser systems, which depend on properties of both the gain medium and the laser cavity, are typically expressed as either double- or single-sided PSDs \cite{darman2017new,ahmed2014theoretical,lu2011experimental,wei2012compact,galzerano2007single,loh2015noise,camatel2008narrow,tran2019tutorial,kim2016ultralow}. 
For consistency, results presented here are in terms of single-sided PSDs. 
All laser systems have similar structural features in their PSDs, with the exact placement and magnitude of these features varying between laser technologies.  Further, free-running relative intensity noise PSD, $\SRIN$, and frequency noise PSD, $\SO$, follow a similar structure (as shown in Figure \ref{fig:laser_noise_models}), and therefore we introduce them together.

The fundamental limits of intensity and frequency noise PSDs take the form of white noise (i.e.\ constant for all Fourier frequencies, $\omega$) \cite{coldren2012diode}. For relative intensity noise, this limit is the shot noise limit (SNL) \cite{obarski2001transfer}
\begin{equation}
    S_\mathrm{RIN}^\mathrm{SNL}(\omega)=\frac{2\hbar\wop}{\bar{P}}\label{eq:SNL}~,
\end{equation}
arising due to the Poisson statistics of the number of photons in the laser beam. Here, $\wop$ is the optical frequency and $\bar{P}$ is the mean power in the optical field. For frequency noise, the white noise floor is given by the quantum noise limit, setting a minimum linewidth of the laser (often called the modified Schawlow–Townes linewidth). The PSD of the quantum noise limit (QNL) takes the value (see Supplementary Information)
\begin{equation}
    S_{\wop}^\mathrm{QNL}(\omega)=\frac{2\pi h\wop\gamma_c^2}{\bar{P}}\label{eq:QNL}~,
\end{equation}
where $\gamma_c$ is the bandwidth of the laser cavity, which is inversely proportional to the cavity length.

In practice, lasers do not typically exhibit fundamentally limited noise PSDs. We refer to all noise processes above the fundamental limit as technical noise. The Fourier frequency response (modulation transfer function) of the gain medium inside a laser cavity acts as a low pass filter for pump noise \cite{coldren2012diode}. The filter bandwidth is approximately set by the inverse of the upper-state lifetime of the laser gain medium, often called the relaxation frequency, $\wrlx$. Below $\wrlx$, the laser has a flat frequency response. Around $\wrlx$, the laser has a resonant response to modulation and undergoes relaxation oscillations. Above $\wrlx$, the relaxation oscillations are damped. The modulation response predominantly transfers pump noise into laser intensity noise. Intensity noise can modulate the refractive index of the gain medium, which alters the phase of the laser light. Therefore, the frequency noise is increased due to the intensity noise, with the increase characterised by the linewidth enhancement factor, $\alpha$ \cite{coldren2012diode}. 

At low Fourier frequencies, the technical noise from the pump takes the form of $1/\omega^a$-type noise (commonly referred to as $1/f$ noise). For low-noise pumps, $1/\omega^a$-type noise is dominated by $1/\omega^1$ (flicker) noise, with higher order $a$ noise being restricted to Fourier frequencies $\omega<2\pi\times 100~$Hz. In all scenarios of interest, we find that these higher order noise terms do not contribute significantly to qubit fidelities unless they span the Rabi frequency of the quantum operation. Therefore, for simplicity, we restrict our study to pure flicker noise.

At the flicker corner frequency, $\wflk$, the flicker noise reaches a white noise floor. For intensity noise, this floor is given by the higher of the pump noise and the SNL. In realistic laser sources, the frequency noise floor for Fourier frequencies from $\wflk$ to $\wrlx$ is enhanced from the QNL by the coupling to intensity noise by a factor of $\alpha^2$ \cite{coldren2012diode}. For Fourier frequencies above the relaxation oscillation peak, pump noise is increasingly damped and the noise amplitude approaches the SNL or QNL accordingly.

In this study we concentrate on two common laser sources used for qubit control: the external cavity diode laser (ECDL) and the diode-pumped solid state laser (DPSSL). A DPSSL typically has higher intensity noise than an ECDL due to pump noise amplification, and lower frequency noise due to a longer laser cavity. ECDLs typically have relaxation oscillation frequencies, $\wrlx$, of order GHz \cite{darman2017new}, whereas for DPSSLs they are sub-MHz \cite{lu2011experimental,wei2012compact,galzerano2007single}.

For ECDLs, the semiconductor's asymmetric gain profile leads to $\alpha$ having typical values between 3 and 6 \cite{giuliani2010linewidth}. For DPSSLs, the gain profile is more symmetric and typically $\alpha\approx0.3$ \cite{fordell2005modulation,thorette2017linewidth} and therefore is not a dominant contribution to frequency noise. Enhanced frequency instability in DPSSLs typically occurs due to mechanical and thermal effects \cite{zhou1985efficient}. 

In addition to the general structural trends of laser noise PSDs, for realistic laser sources there are bumps and spurs present in the laser spectrum. These features typically occur due to mechanical instabilities and actuator resonances in the laser cavity. For well designed laser cavities (e.g. Toptica DL Pro \cite{topticaDLpro}), relatively smooth laser noise PSDs can be achieved. In this study, we focus on the gross structural features of laser noise PSDs due to the lasing process itself. Uncontrolled spurs in laser noise PSDs will affect qubit fidelities, however their influence is limited to when the Rabi frequency is close to the spur frequency. Calculations of qubit control fidelities for a measured laser frequency noise PSD is presented in the Supplementary Information.

The technical noise present in both intensity and frequency noise of laser sources often requires the stabilisation of laser sources for use in quantum control. We model the PSDs of stabilised laser sources with the simplified model
\begin{equation}
\Ssrv =
  \begin{dcases}
                                   h_a & \text{for $\omega<\wsrv$} \\
  h_b & \text{for $\omega>\wsrv$}
  \end{dcases} ~,
  \label{eq:simpleservo}
\end{equation}
such that $h_a$ is a constant white noise amplitude within the stabilisation bandwidth, $\wsrv$, and $h_b$ is the residual free-running white noise amplitude outside the stabilisation bandwidth. We assume that $h_a<h_b$, and we address the case where this is not automatically valid (e.g. solid state lasers) in the Discussion. Such a model has previously been used to explore the effect of stabilisation bandwidths on laser linewidth reduction \cite{di2010simple}. In the followings section we connect this general PSD to qubit fidelities and derive requirements on laser frequency and intensity noise to achieve high fidelity qubit control.

\subsection{Frequency noise stabilisation requirements}

\begin{figure*}
    \centering
    \includegraphics[width=0.75\linewidth]{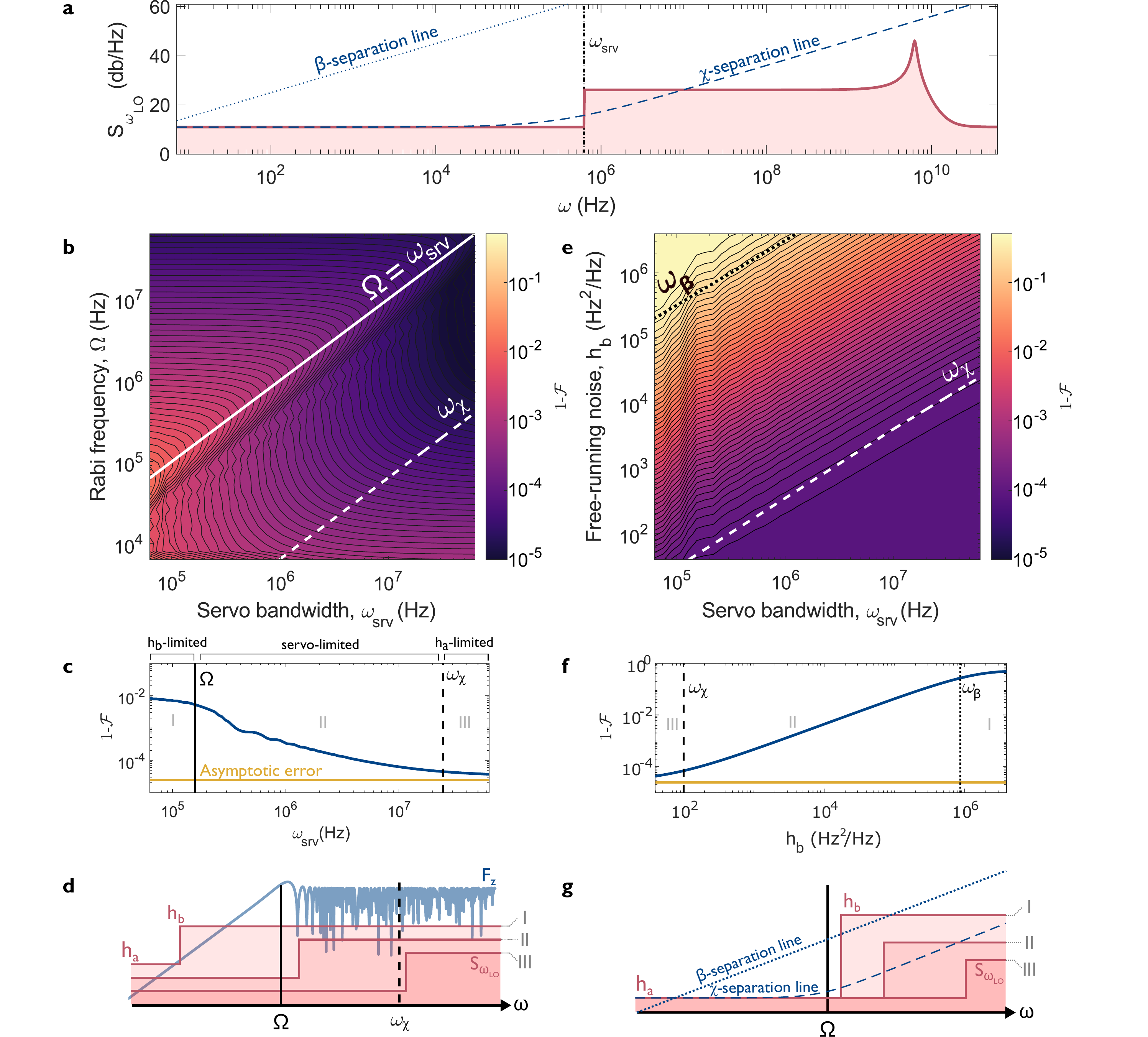}
    \caption{\textbf{The infidelity parameter space for a servoed diode laser with a $\bm{\Gamma_\mathrm{FWHM}=1~}$Hz narrowed linewidth ($\bm{h_a=4\pi\times 1~}$Hz$^2$/Hz)}, (a) The frequency noise PSD of a LO generated from a servoed diode laser. Within the servo loop, the noise amplitude takes the value of $h_a$. Above the servo bandwidth, $\wsrv$, the LO takes on the free-running laser noise including a relaxation oscillation peak of $20~$dB at $2\pi\times1~$GHz. (b) The infidelity landscape as the Rabi frequency and servo bandwidth are varied, demonstrating three main regions (detailed in text). The $h_b$-limited region is separated by $\Omega=\wsrv$ (solid line) from the servo-limited region region, with the $\chi$-separation line (dashed) delineating the $h_a$-limited region. (c) A 1D cross-section of (b) at $\Omega=\pi/2\times10^5~$Hz. (d) A diagrammatic representation of noise spectral densities in each of the three regions, with the $\pi-$pulse filter function (in arbitrary units) overlaid for comparison. (e) The infidelity landscape as the above-servo noise amplitude, $h_b$, and the servo bandwidth are varied for $\Omega=2\pi\times10~$kHz. The $\beta$-separation line (dotted), does not guarantee qubit coherence, while the $\chi$-separation line (dashed) again bounds the $h_a$-limited region. (f) A cross-section of (e) at $\wsrv=300~$kHz. (f) A diagrammatic representation of the PSDs in each of the three regions with respect to the $\beta$- and $\chi$-separation lines. The $\chi$-separation line in Fourier space is given by Equation \ref{eq:xi_sep_line}.
    }
    \label{fig:fidelity_servo_laser}
\end{figure*}

For LOs derived from laser radiation, the frequency noise of the LO directly results from noise in the laser source. 
In this instance the PSD of LO frequency noise, $\SLO$, is equivalent to the stabilised laser frequency noise, which we approximate with the simple expression in Equation \ref{eq:simpleservo}. 
For the purposes of a general discussion on how stabilised frequency noise affects qubit control, we express all results in terms of the parameters $h_a$, $h_b$ and $\wsrv$. 
For an exposition of the specific way that laser noise processes connect to these parameters for different qubit and laser types, see the Supplementary Information.

To connect laser noise parameters to qubit fidelities for arbitrary single-qubit operations, we require a general expression for single-qubit filter functions. 
We approximate a general filter function of a quantum operation as the piece-wise function 
\begin{equation}
    F^{(u)}(\omega) =
  \begin{dcases}
  c_a^{(u)}\omega^{n_u} & \text{for $\omega<\wcut^{(u)}$} \\
  c_b^{(u)} & \text{for $\omega>\wcut^{(u)}$}
  \end{dcases}~,
  \label{eq:generalFF}
\end{equation}
such that the qubit has a flat response to control noise above the cut-off frequency, $\wcut^{(u)}$, and noise below the cut-off is damped at $10\log_{10}(n_u)$ dB per decade, where $n_u$ is the order of the filter function. 
The condition of continuity sets the requirement $c_b^{(u)}=c_a^{(u)}(\wcut^{(u)})^{n_u}$, and $c_{a,b}$ are constants set by the form of the filter function. 
Such a general model can approximate a wide range of quantum operations, including composite pulse sequences designed to correct for control errors and noise \cite{ball2015walsh,bermudez2017assessing}.

We find that to improve qubit control over using a free-running laser, the servo bandwidth must be above the filter function cut-off frequency, which in the case of a primitive $\pi-$pulse is the Rabi frequency. Substituting the piece-wise expressions for the frequency noise PSD (Equation \ref{eq:simpleservo}) and the general filter function into Equation \ref{eq:FF_fidelity_maintext} (see Supplementary Information), we can derive an approximate expression for the fidelity decay constant, $\chi^{(u)}$, of a general single-qubit operation. In the regime that the servo bandwidth is below the filter function cut-off frequency, $\wsrv<\wcut$, we find $\chi^{(u)}\approx n c^{(u)}_b h_b/(4\pi(n-1)\wcut)$. 
In this instance, the qubit fidelity is dominated by the free-running noise of the laser, $h_b$, and the contribution from $h_a$ has negligible influence on qubit errors. We refer to this regime as \emph{$h_b$-limited}.

In the regime $\wsrv>\wcut$, the expression for the fidelity decay constant becomes
\begin{equation}
\chi^{(u)}=\frac{c^{(u)}_b}{4\pi}\left( \frac{n_u}{n_u-1} \frac{h_a}{\wcut^{(u)}}  +  \frac{h_b-h_a}{\wsrv}\right)~.
\label{eq:fidelity_servo}
\end{equation}
The first term contains the ratio of the stabilised frequency noise $h_a$ to the cut-off frequency of the quantum operation, $\wcut^{(u)}$, which determines a fundamental limit to the fidelity based on the stabilised laser noise. 
The second term places a limit on the fidelity from the free-running noise, $h_b$, and the servo bandwidth, $\wsrv$. 
For servo bandwidths
\begin{equation}
    \wsrv>\omega_{\chi}^{(u)} \equiv\frac{(n_u-1)\wcut^{(u)}}{n_u}\left(\frac{h_b}{h_a}-1\right)~,
    \label{eq:xi_line_servo}
\end{equation}
 the first term in Equation \ref{eq:fidelity_servo} exceeds the second term such that $h_a$ is the dominant contribution to the qubit fidelity (\emph{$h_a$-limited}). For servo bandwidths below this frequency, the fidelity is limited by the insufficient suppression of the free-running noise (\emph{servo-limited}). Here, $\omega_{\chi}^{(u)}$ defines the cutoff between the $h_a$-limited and servo-limited regions.


As an example, we consider the filter function for the primitive $\pi$-pulse with Rabi frequency $\Omega$. In this instance, $c_b^{(\pi)}=4$, $n_\pi=2$ and $\wcut^{(\pi)}=\Omega$ (see Methods), such that Equation \ref{eq:fidelity_servo} becomes
\begin{equation}
\chi^{(\pi)}=\frac{1}{\pi}\left( \frac{2h_a}{\Omega}  +  \frac{h_b-h_a}{\wsrv}\right)~,
\label{eq:fidelity_servo_pi}
\end{equation}
and Equation \ref{eq:xi_line_servo} becomes
\begin{equation}
    \wsrv> \omega_{\chi}^{(\pi)} =\frac{\Omega}{2}\left(\frac{h_b}{h_a}-1\right)~.
    \label{eq:xi_line_servo_pi}
\end{equation}


Equation \ref{eq:xi_line_servo_pi} shows that the $\pi$-pulse error from laser noise is dominated by residual noise in the region
\begin{equation}
    \SLO\ge h_a\left(\frac{2\omega}{\Omega}+1\right)~. \label{eq:xi_sep_line}
\end{equation} 
We define the boundary of this region as the \emph{$\chi$-separation line}. The requirement on the servo bandwidth is therefore to suppress all free-running noise to below the $\chi$-separation line. 
 For typical values of $h_a$ and $\Omega$, this limitation is stricter than the requirement on narrowing the linewidth of a laser source from the $\beta$-separation line (see Supplementary Information), 
\begin{equation}
    \SLO\ge \pi\omega~.\label{eq:beta_sep_line}
\end{equation} 
 The $\beta$-separation line divides the frequency noise PSD into a region that contributes to the laser linewidth and a region that only contributes to the lineshape wings (see Figure \ref{fig:laser_noise_models}(b)). The required servo bandwidth to narrow the linewidth of the laser is the one that suppresses all free-running noise to below this $\beta$-separation line, while a much higher servo bandwidth is needed to suppress noise below the $\chi$-separation line to minimise $\pi$-pulse control errors.


To illustrate the above findings and confirm that the $\chi$-separation line provides a useful measure of fidelity optimization, we show numerical calculations of primitive qubit infidelities in Figure \ref{fig:fidelity_servo_laser}. 
We use the the frequency noise PSD shown in Figure \ref{fig:fidelity_servo_laser}(a) and the exact expression for the first-order filter function of a $\pi-$pulse (see Methods, Equation \ref{eq:prim_filter}). 
Such a PSD applies equally to a single servoed ECDL addressing an optical qubit, or two phase-locked ECDLs addressing a hyperfine qubit. 
In contrast to the simplified-model PSD used to derive Equation \ref{eq:fidelity_servo}, we have included the relaxation oscillation peak found in ECDLs.


 Our simulations confirm the existence of the three regions -- $h_b$-limited, servo-limited, and $h_a$-limited -- that were identified in our analysis using piece-wise approximations.
 These regions can be seen in Figure \ref{fig:fidelity_servo_laser}(b) and (c) as the servo bandwidth, $\wsrv$ is varied. 
For $\wsrv<\Omega$ (the $h_b$-limited region), changing the servo bandwidth has little effect on the fidelity. 
The reason for this is illustrated by Figure \ref{fig:fidelity_servo_laser}(d), where the filter function in frequency space is shown for a primitive operation. 
The response of the filter function is flat for Fourier frequencies above $\Omega$, and free-running noise in this region dominates the qubit errors. 
For $\Omega<\wsrv<\omega_{\chi}^{(\pi)}$ (the servo-limited region), the contribution of $h_b$ is reduced by the stabilisation loop; however, it can still have a sizeable effect compared to the contribution to the error from $h_a$. 
To improve fidelity in this regime, the servo bandwidth must be increased, as the fidelity is largely independent of $\Omega$. 
 For $\wsrv>\omega_\chi^{(\pi)}$ (the $h_a$-limited region), the contribution to the fidelity from $h_a$ becomes dominant, as $h_b$ is suppressed below the $\chi$-separation line and increasing the servo bandwidth further provides diminishing returns.


We also confirm numerically the importance of the $\chi$-separation line to qubit fidelities.
 In Figure \ref{fig:fidelity_servo_laser}(e) and (f) the $\beta$- and  $\chi$-separation lines are directly compared. 
For high values of $h_b$ and low servo bandwidths, there is no coherence between the laser source and the qubit, indicated by fidelities of $\mathcal{F}=0.5$. 
The $\beta$-separation line relates only to properties of the laser, and therefore does not necessarily imply qubit coherence, as seen in the case shown in Figure \ref{fig:fidelity_servo_laser}(e) and (f).
As $h_b$ decreases and/or $\wsrv$ increases, fidelities improve up until the $\chi$-separation line where the fidelity becomes limited by $h_a$. 
These trends are further illustrated in Figure \ref{fig:fidelity_servo_laser}(g). 
In the instance that any part of the free-running noise is above the $\beta$-separation line, this free-running noise dominates the FWHM linewidth.
Similarly, when any part of the free-running noise is above the $\chi$-separation line, that free-running noise reduces the fidelity compared with an idealized white-noise LO of the same FWHM linewidth.

\begin{figure}
    \centering
    \includegraphics[width=1\linewidth]{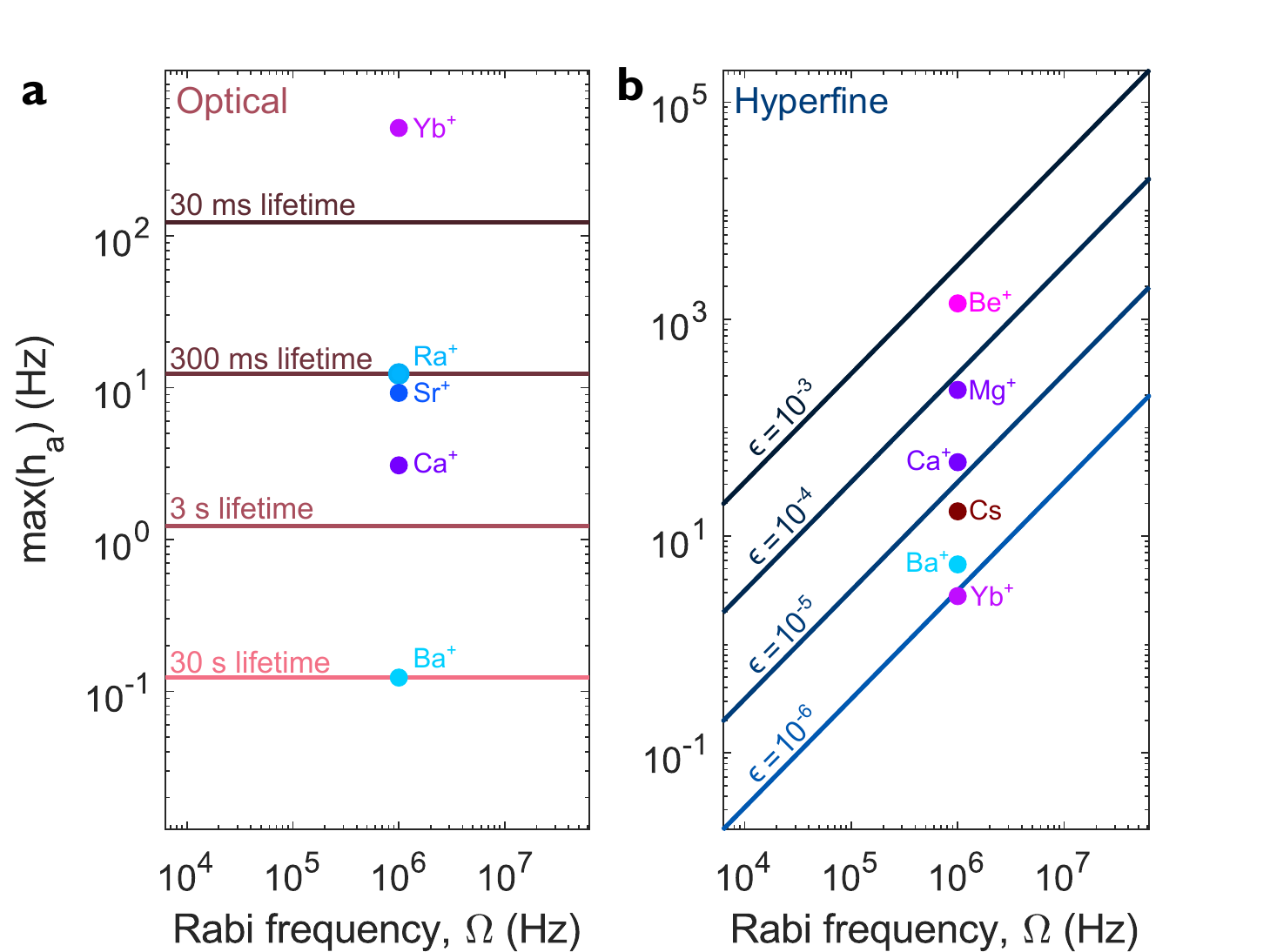}
    \caption{\textbf{Maximum allowed laser frequency noise to be at the SE floor in the infinite servo bandwidth limit for a primitive operation}. In this limit the linewidth of the LO is approximately equivalent to $\Gamma_\mathrm{FWHM}=h_a/4\pi$.  (a) Optical qubits, where the SE floor is given by the lifetime of the qubit state. Typical lifetimes of metastable atomic states are used, which approximately correspond to SE errors, $\epsilon_\mathrm{SE}$, of $6\times10^{-6}$, $6\times10^{-5}$, $6\times10^{-5}$, and $6\times10^{-3}$, from longest to shortest respectively. The SE floors of $S_{1/2}\rightarrow D_{5/2}$ transitions in a selection of optical qubit candidates are shown for reference \cite{UDportal,bruzewicz2019trapped}. (b) Hyperfine qubits, where the SE floor is given by off-resonant scattering. We define $\epsilon=1-\mathcal{F}$. The requirement on laser noise is relaxed for higher Rabi frequency as less noise is sampled by the qubit and the SE error is constant with Rabi frequency. The value of $h_a$ corresponding to the minimum value of $\epsilon_\mathrm{SE}$ for a selection of hyperfine qubit candidates at $\Omega=1~$MHz are shown for reference.}
    \label{fig:fidelity_limits}
\end{figure}


To determine the conditions under which the error from an $h_a$-limited laser source is smaller than the fundamental SE error, we set the requirement $\epsilon_\mathrm{SE}>\chi^{(u)}/2$. Here, $\epsilon_\mathrm{SE}$ is the spontaneous emission error (see Methods for definition). In the asymptotic limit, $\wsrv\rightarrow\infty$, this rearranges to the requirement
\begin{equation}
h_a<\frac{8\pi(n_u-1)\epsilon_\mathrm{SE}\wcut^{(u)}}{n_u c_b^{(u)}}~,
\end{equation}
which for a $\pi$-pulse reduces to $h_a<\pi\epsilon_\mathrm{SE}\Omega$. For a finite value of $\wsrv$, the fidelity does not appreciably change from this asymptote in the $h_a$-limited region, as seen in Fig. \ref{fig:fidelity_servo_laser}(c).
The maximum value of $h_a$ that saturates this bound is plotted in Figure \ref{fig:fidelity_limits} for both optical and hyperfine qubits with varying spontaneous emission floors. 
For optical qubits, the spontaneous emission error is inversely proportional to the Rabi frequency (see Methods), setting a constant requirement on $h_a$ for all $\Omega$. For hyperfine qubits, the spontaneous emission error is constant with Rabi frequency; thus, increasing $\Omega$ allows higher values of $h_a$ to satisfy the inequality in Equation (12).

\subsection{Intensity noise}



The results presented for the effect of laser frequency noise on qubit control share many similarities with the effect of laser intensity noise. For stabilised intensity noise, we approximate the PSD, $\SRIN$, using the simple model of Equation \ref{eq:simpleservo}, with noise amplitudes $h_a'$ and $h_b'$ replacing $h_a$ and $h_b$, respectively. Here, $h_a'$ and $h_b'$ have different physical units than $h_a$ and $h_b$.
Similar to our analysis for frequency noise, the fidelity decay constant, $\chi^{(u)}$, for intensity noise takes the form
\begin{equation}
\chi^{(u)}=\frac{\kappa\Omega^2 c^{(u)}_b}{4\pi}\left( \frac{n^{(u)}}{n^{(u)}-1} \frac{h_a'}{\wcut^{(u)}}  +  \frac{h_b'-h_a'}{\wsrv}\right)~,
\label{eq:fidelity_servo_intensity}
\end{equation}
with the constant $\kappa=1/4$ for optical qubits and $\kappa=1$ for hyperfine qubits.
The additional factor of $\kappa\Omega^2$ over the fidelity decay constant for frequency noise in Equation \ref{eq:fidelity_servo} comes from the conversion of intensity noise to Rabi frequency noise (see Methods). 
Thus, the servo bandwidth requirement and the $\chi$-separation line take the same form as for frequency noise in Equations \ref{eq:xi_line_servo} and \ref{eq:xi_sep_line} respectively. 

Similarly to frequency noise, we derive the requirement for the intensity noise error to be below the fundamental SE floor by enforcing the constraint $\epsilon_\mathrm{SE}>\chi^{(u)}/2$. In the asymptotic limit, $\wsrv\rightarrow\infty$, this becomes
\begin{equation}
h_a'<\frac{8\pi(n-1)\epsilon_\mathrm{SE}\wcut^{(u)}}{n \kappa \Omega^2 c_b^{(u)}}~,\label{eq:ha_limit_RIN}
\end{equation}
which simplifies to $h_a'<2\epsilon_\mathrm{SE}/(\kappa\Omega)$ for a $\pi$-pulse. 


The fundamental limit of laser intensity noise, and therefore $h_a'$, is the SNL (Equation \ref{eq:SNL}). Therefore, the lowest achievable error depends on the SNL, which is inversely proportional to laser power. As $\Omega$ increases with increasing laser intensity, the condition of Equation \ref{eq:ha_limit_RIN} can be reduced to involve only the beam size, which is determined by the usable numerical aperture (NA) of the addressing beam. We find (see Supplementary Information) that for both hyperfine and optical qubits, the condition is satisfied for all physical numerical apertures up to the Abbe limit in vacuum (NA=1). Therefore, the intensity noise error from SNL laser light is always below the SE floor. This result is universally true across all single-valence electron atoms (for hyperfine qubits) and quadrupole transitions (for optical qubits), as the atomic and laser parameters that change from system to system are contained within $\epsilon_\mathrm{SE}$.

As well as contributing to Rabi noise, the intensity noise can couple to an effective dephasing noise through AC Stark shifts. 
The varying laser intensity changes the effective electric field strength at the atomic position, leading to a time-varying change in the qubit frequency. 
Assuming a static laser frequency, this causes an effective detuning that can lead to a non-trivial error contribution from dephasing. We neglect the contribution of AC Stark shift for optical qubits driven on resonance, and the following results apply to hyperfine qubits. 

For CW laser radiation, the Stark shift, $\Delta^\mathrm{(cw)}_\mathrm{AC}$, depends on the two-photon Rabi frequency, $\Omega_{2\gamma}$, as $\Delta^\mathrm{(cw)}_\mathrm{AC}=\mu_\mathrm{cw} \Omega_{2\gamma}$. Here, the dimensionless proportionality constant, $\mu_\mathrm{cw}$, can be calculated from atomic and laser parameters and its value typically takes an order of magnitude of $10^{-3}$ \cite{wineland2003quantum}. 
We find that Stark shift noise from CW laser radiation causes less infidelity than Rabi noise for $\mu_\mathrm{cw}<\sqrt{(\pi^2+4)/8}\approx1.3$ (see Supplementary Information), and is therefore typically negligible.

When the laser radiation is a frequency comb, each comb tooth frequency contributes to the Stark shift, $\Delta^\mathrm{(fc)}_\mathrm{AC}$, and its magnitude depends quadratically on the two-photon Rabi frequency as $\Delta^\mathrm{(fc)}_\mathrm{AC}=\mu_\mathrm{fc} \Omega_{2\gamma}^2$. The proportionality constant, $\mu_\mathrm{fc}$, is again calculated from atomic and laser parameters, typically taking the value of $\sim 10^{-9}~$Hz$^{-1}$ (see Supplementary Information). We find that infidelities from frequency comb Stark shifts are below those of Rabi noise for $\mu_\mathrm{fc}<\sqrt{(\pi^2+4)/(32\Omega_{2\gamma}^2)}\approx 0.65/(\Omega_{2\gamma})$. Therefore, for typical Stark shift values and Rabi frequencies, AC Stark shift noise from intensity noise is not a dominant contribution.

In the above analysis we have concentrated on the noise from the lasing process itself. There are other sources of intensity noise between the laser head and the atom position, such as AOM diffraction efficiency noise, polarisation-to-intensity noise and beam pointing jitter. These effects may contribute a greater source of intensity noise than the laser itself, especially for tightly focused individual addressing beams. However we consider these technical, and not fundamental, limits to qubit fidelities and their consideration is beyond the scope of our present analysis.

\section{Discussion}

We have outlined the requirements for laser sources to perform quantum control with errors below the SE floor. The noise of a free-running laser necessitates the use of stabilisation loops to reach this fundamental floor, and we have shown how the specific details of this loop are important for fidelity optimisation. Specifically, there is an interplay between the stabilized noise amplitude, $h_a$, the servo bandwidth, $\wsrv$, the Rabi frequency, $\Omega$, and the residual free-running noise of the noise spectrum, $h_b$. The interplay can be summarised in the introduced concept of the $\chi$-separation line, such that when free-running noise is suppressed below this line, the fidelity of qubit operations will be approximately limited at a fundamental level by the value of $h_a$. Within this $h_a$-limited regime, noise from the laser source can become non-dominant by appropriately minimising the value of $h_a$, which can be below the SE floor. Therefore, the task of performing optimal quantum control using laser LOs is first to determine the required value of $h_a$ such that errors are below the SE floor, and then to determine the appropriate servo bandwidth to suppress free-running noise such that $h_a$-limited operation is achieved.

The simple PSD used to model stabilised laser sources does not capture some non-universal features of realistic laser noise spectra, such as spurs and servo bumps. However, we find numerically that deviations from the simple model do not compromise the use of the $\chi$-separation line as a metric for laser noise optimisation. For example, when we introduce a servo bump at $\wsrv$, we find that if the noise is below the $\chi$-separation line, its influence is negligible. Similarly, for the relaxation oscillation peak we numerically find that in the regime where the $\wrlx$ is close to $\wsrv$, the relaxation peak has a negligible influence as long as it sits below the $\chi$-separation line. These examples suggest that any noise below the $\chi$-separation line has little influence on qubit fidelities, irrespective of its exact structure. This conclusion is further supported by the application of the $\chi$-separation line to the frequency noise PSD of an ultra-stable laser from Menlo Systems (see Supplementary Information), where the $\chi$-separation line correctly predicts the numerically calculated $\pi-$pulse error despite spurs and servo bumps being present.

Our results on the influence of LO noise from a laser source perspective also provide informative design constraints on LOs derived from microwave sources. It has previously been shown that lab-grade microwave oscillators can cause significant fidelity limitations on qubit operations, and that composite pulse sequences provide a negligible improvement in achievable fidelities \cite{ball2016role}. These lab-grade oscillators have a remarkably similar frequency noise PSD to a diode laser locked to a high finesse cavity ($h_a\sim10^{-1}~$Hz), with a servo bandwidth of approximately $2\pi\times10^4~$Hz. As we have shown, such a servo bandwidth is insufficient to narrow the effective linewidth the qubit experiences. Similarly, it was shown that a precision LO has a smaller low-frequency noise ($h_a\sim10^{-4}~$Hz), but is still servo-limited in its operation. Therefore, to improve the use of microwave oscillators for driving qubit operations, either the phase-locked-loop bandwidth must be increased, or the intrinsic phase noise of the variable oscillator must be improved. For microwave oscillators with $\Omega=100~$kHz, servo bandwidths would have to be increased to approximately $5~$MHz to achieve $h_a$-limited operation.

The simple PSD model used in this manuscript does not match the typical noise spectra of stabilised DPSSLs. Solid state gain media typically have long relaxation times, and therefore low values of $\wrlx$ (of order $2\pi\times10^5~$kHz), and it is possible for $\wsrv<\wrlx$ such that the relaxation peak is suppressed. In this instance the free running frequency noise is actually the QNL, such that $h_a>h_b$. In this case, the qubit fidelities are automatically limited by the value of $h_a$ without the servo bandwidth having to satisfy the requirement from the $\chi$-separation line. Therefore, DPSSLs have a distinct advantage over ECDLs in that $h_a$-limited operation can be achieved with comparably relaxed servo bandwidth requirements.

In the instance where active stabilisation bandwidths needed to achieve $h_a$-limited fidelities are technologically demanding, or even prohibitive, it may be preferable to use passive stabilisation techniques. For an LO of an optical qubit, this can be performed using the transmitted light of a high-finesse cavity, such that the resulting laser frequency noise is low-pass filtered by the cavity linewidth \cite{fluhmann2015spectral}. Offset injection locking can be used to passively lock two diode lasers, where the light of a primary laser is shifted in frequency by the qubit frequency and injected into a secondary laser \cite{linke2013injection}. The effective bandwidth of the phase locking is that of the cavity bandwidth, which, for the short cavity lengths present in ECDLs, can be of order GHz. Similar to frequency noise, the suppression of intensity noise could be performed passively to avoid the demanding requirements on servo bandwidths. The use of a saturated optical amplifier can generate SNL light with watts of optical power over at least a $50~$MHz bandwidth\cite{yang2017high}. Alternatively, collinear balanced detection can be used as a notch filter on laser intensity noise, with center frequencies of the notch filter being able to be passively selected from MHz to GHz \cite{allen2019passive}. Therefore, this technique can be used to suppress noise around $\wcut$ .

The interplay between laser noise and an arbitrary filter function was derived. In the general expressions of Equations \ref{eq:fidelity_servo} and \ref{eq:xi_line_servo}, the order of the filter function, $n_u$, plays an important role in both the achievable fidelity and the required servo bandwidth. As $n_u$ is increased (better low frequency filtering), the fidelity is improved. However, filter functions of higher order $n_u$ are constructed using concatenated pulses. For the same laser power, these concatenated sequences are longer in duration than a primitive $\pi$-pulse, reducing $\wcut$. Therefore, there is a competing effect between the increasing order of $n_u$ and decreasing value of $\wcut$. This result is reflected in the previous finding of an unreliable improvement in fidelity of a dynamically corrected gate (DCG) for typical microwave oscillators that have approximately the same frequency PSD as Equation \ref{eq:simpleservo} \cite{ball2016role}. We have confirmed these findings for the simple model PSD using a fixed laser power, and find negligible improvement in fidelity of a DCG over a $\pi-$pulse. A slight improvement is only found in the instance that noise is suppressed below the $\chi$-separation line. Further investigations are required to explore whether these improvements are maintained for actual laser noise PSDs.

The analysis presented here has focused on single-qubit gates, where the sideband noise causes infidelities on the qubit carrier transition. In physical atomic systems of interest, there are often other nearby transitions that the sideband noise could interact with, such as Zeeman states, or motional modes due to confinement. The coupling of sideband noise to these transitions could lead to additional infidelity pathways not quantified here. In particular, when the motional modes are used for two-qubit entangling operations, such as the M{\o}lmer-Sorensen gate in trapped ion chains, excess phase noise at the carrier transition frequency will cause two-qubit gate infidelities. The analysis here should therefore be extended to Hamiltonians beyond the two-level qubit to include additional transitions and quantum operations. These further analyses would require extensions to the filter function theory formalism employed here. These extensions would allow for the studies to be applied more broadly to other fields, such as quantum simulation experiments, with the Ising Hamiltonian as one example. Such analyses would allow for the appropriate choice and tailoring of LOs for a vast array of quantum systems of interest, to maximise the fidelity and usefulness of these apparatus.

\section{Methods}

\subsection{Spontaneous emission}

For optical qubits, the fundamental limit to qubit fidelities is from the finite upper-state lifetime.The fidelity of a primitive pulse to an excited state of finite lifetime,$\tau_e$, is given by \cite{loudon2000quantum}
\begin{equation}
    \mathcal{F}=1-\epsilon_\mathrm{SE}=\frac{1}{2}\left(1+ e^{-\frac{3\pi}{4\Omega\tau_e}}\right)~,
    \label{eq:SE_optical}
\end{equation}
such that  higher Rabi frequency and longer state lifetimes leads to higher gate fidelities.

For microwave qubits driven by Raman transitions, the spontaneous emission is due to photons off-resonantly scattering from the intermediate state. For equal intensity laser beams driving $\sigma^\pm$ transitions, the scattering probability is independent of the Rabi frequency, such that \cite{campbell2010ultrafast}
\begin{equation}
\mathcal{F}=1-\epsilon_\mathrm{SE}=1-\Gamma\pi\left(\frac{3}{\omega_F} - \frac{1}{\Delta} + \frac{2}{\Delta-\omega_F}\right) ~,
\label{eq:SE_hyperfine}
\end{equation}
where $\Gamma$ is the linewidth of the intermediate transition, $\omega_F$ is the splitting to the next closest excited state to the intermediate transition, and $\Delta$ is the detuning of the laser light from the intermediate transition. The off-resonant scattering probability can be minimised for $\Delta\approx0.4\omega_F$, and its exact value is strongly dependent on the atomic species used, due to different values of $\Gamma$ and $\omega_F$ that can vary over orders of magnitude. Typical values of the SE error are presented in the Supplementary Information.

\subsection{The Hamiltonian}

We consider the Hamiltonian of an atom interacting with a LO field and subject to time-dependent errors in its frequency and Rabi frequency \cite{green2013arbitrary,kabytayev2014robustness,ball2016role},
\begin{equation}
\mathcal{H}(t)=\frac{1}{2} \delta\Delta(t)\sigma_z + \frac{1}{2}(\Omega_c(t)+\delta\Omega(t))\sigma_{\theta}~,\label{eq:hamiltonian}
\end{equation}
where $\sigma_{\theta}=\{\cos[\phi_c(t)]\sigma_x+\sin[\phi_c(t)]\sigma_y\}$, $\Omega_c(t)$ and $\phi_c(t)$ are the Rabi frequency and phase of the control field respectively, $\delta\Delta(t)$ describes the time-varying detuning of the LO frequency from the qubit frequency, and $\delta\Omega(t)$ is the  time-varying fluctuations of the Rabi frequency. The Hamiltonian can be used to represent both optical and two-photon Raman transitions.

For optical transitions $\Omega_c\propto\sqrt{I}$, where $I$ is the laser intensity, the relationship between RIN and Rabi frequency noise is given by
\begin{equation}
    \delta\Omega(t)=\frac{1}{2}\Omega_c\frac{\delta I(t)}{I} ~,\label{eq:onephotonrabi}
\end{equation}
where $\delta I(t)$ is the time-varying intensity noise of the laser field. 

For two-photon Raman transitions $\Omega\propto \sqrt{I_1}\sqrt{I_2}$, where $I_1$ and $I_2$ are the intensities of each laser beam. Making the assumption that $I_1=I_2=I$, and the Rabi frequency noise is then related to the RIN by
\begin{equation}
    \delta\Omega(t)=\Omega_c\frac{\delta I(t)}{I}~.
\end{equation}
In the instance of the detuning of the LO field from the qubit resonance being entirely due to LO frequency noise, $\delta\Delta=\delta\omega_{LO}$. 

For CW Raman transitions, the intensity-to-frequency conversion of AC Stark shifts enters the Hamiltonian as an effective frequency detuning term, $\Delta^\mathrm{(cw)}_\mathrm{AC}\propto g^2$, where $g\propto\sqrt{I}$ is the contribution of one laser beam's intensity to the two-photon Rabi frequency. As $\Omega_{2\gamma}\propto I$, variations in the AC Stark shift with intensity noise lead to a noise field of
\begin{equation}
\delta\Delta^\mathrm{(cw)}_\mathrm{AC}= \mu_\mathrm{cw} \Omega_{2\gamma} \frac{\delta I}{I} ~,
\end{equation}
where $\Delta^\mathrm{(cw)}_\mathrm{AC}=\mu_\mathrm{cw} \Omega_{2\gamma}$. Similarly, for a ML laser, where all comb lines contribute to the AC Stark shift such that
\begin{equation}
\delta\Delta^\mathrm{(fc)}_{z,\mathrm{AC}}=2\mu_\mathrm{fc} (\Omega'_{2\gamma})^2 \frac{\delta I}{I} ~,\label{eq:fcstark}
\end{equation}
where $\Delta^\mathrm{(fc)}_\mathrm{AC}=\mu_\mathrm{fc} (\Omega'_{2\gamma})^2$. Typical values of $\mu_\mathrm{cw}$ and $\mu_\mathrm{fc}$ are presented in the Supplementary Information.

\subsection{Filter function theory}

Consider a classical noise field $\beta_j(t)$ such that it contributes an error term to the Hamiltonian
\begin{equation}
    \mathcal{H}_\mathrm{err}(t)=\beta_j(t)\sigma_j~,
\end{equation}
where $j=\{z,\theta\}$. The single-sided PSD of $\beta_j(t)$ is defined through the Weiner-Khinchin theorem as
\begin{equation}
\langle\beta_j(t_1)\beta_j(t_2)\rangle=\frac{1}{2\pi}\int^\infty_{0} S_j(\omega)e^{i\omega(t_2-t_1)}d\omega~,
\end{equation} 
where $\langle\cdot\rangle$ denotes ensemble-averaging. The total fidelity of a unitary operation, defined by its filter function $F^{(u)}_j(\omega)$ can be calculated to first order as \cite{ball2015walsh,kabytayev2014robustness,green2013arbitrary}
\begin{equation}
 \mathcal{F}^{(u)}(\tau)=\frac{1}{2}\left\{1+\exp[-\chi^{(u)}(\tau)]\right\}  ~, \label{eq:FF_fidelity}
\end{equation}
with
\begin{equation}
    \chi^{(u)}(\tau)=\sum_j \frac{1}{\pi}\int_0^\infty \frac{d\omega}{\omega^2} S_j(\omega) F_j^{(u)}(\omega)~,
\end{equation}
such that $S_z$ and $S_\theta$ determine the dephasing and amplitude noise errors. These PSDs can be related to physical noise processes by again considering the Weiner-Khinchin theorem. Expressing the noise field as $\beta_j=\alpha f_j(t)$, where $\alpha$ is constant and $f_j(t)$ is a time-varying function. Through the linearity of the ensemble-averaging operation
\begin{align}
\langle\beta_j(t_1)\beta_j(t_2)\rangle&=\alpha^2\langle\delta f_j(t_1)\delta f_j(t_2)\rangle ~,\\
&=\frac{\alpha^2}{2\pi}\int^\infty_{0} S_{f_j}(\omega)e^{i\omega(t_2-t_1)}d\omega~,
\end{align}
such that $S_j(\omega)=\alpha^2 S_{f_j}(\omega)$, providing a way to convert between PSDs given their functional relation of the corresponding parameters. 

Inspecting the full Hamiltonian (Equation \ref{eq:hamiltonian}) the noise fields are given by
\begin{align}
    \beta_z(t)&=\frac{1}{2}\delta\Delta(t)~, \\
    \beta_\theta(t)&=\frac{1}{2}\delta\Omega(t)~,
\end{align}
such that the relation between the PSDs of these parameters are
\begin{align}
    S_z(\omega)&=\frac{1}{4}S_{\Delta}(\omega)~, \\
    S_\theta(\omega)&=\frac{1}{4}S_{\Omega}(\omega)~.
\end{align}
The PSDs for dephasing and amplitude errors can then be related to physical noise processes using the expressions in Equations (\ref{eq:onephotonrabi})-(\ref{eq:fcstark}). For example, for optical transitions,  $S_{\Omega}(\omega)=\frac{1}{4}\Omega_c^2 S_\mathrm{RIN}(\omega)$ such that
\begin{equation}
    S_\theta(\omega)=\frac{1}{16}\Omega_c^2 S_\mathrm{RIN}(\omega)~,
\end{equation}
and for LO frequency noise $S_{\Delta}(\omega)=S_{\omega_\mathrm{LO}}(\omega)$, such that
\begin{equation}
    S_z(\omega)=\frac{1}{4} S_{\omega_\mathrm{LO}}(\omega) ~,
\end{equation}
which is the expression previously used in Ref. \cite{ball2016role}.

Once the PSDs are well defined, the fidelity can be calculated using knowledge of the corresponding filter functions, $F_z^{(u)}(\omega)$ and $F_\theta^{(u)}(\omega)$, for the desired operation. For a $\pi$-pulse around $X$, the dephasing filter function is given by
\begin{equation}
F^{(\pi)}_z(\omega)=\left\lvert \frac{\omega^2}{\omega^2 - \Omega^2} (e^{i\omega\tau_\pi} + 1)\right\rvert^2+\left\lvert \frac{i\omega\Omega}{\omega^2 - \Omega^2} (e^{i\omega\tau_\pi} + 1)\right\rvert^2~,
\label{eq:prim_filter}
\end{equation}
where $\tau_\pi=\pi/\Omega$. By Taylor expanding $F^{(\pi)}_z(\omega)$ and taking the leading order, $F^{(\pi)}_z(\omega)$ can be approximated by the piecewise function
\begin{equation}
    F^{(\pi)}_{z}(\omega)\approx
  \begin{cases}
  4\omega^2/\Omega^2 & \text{for $\omega<\Omega$}\\
  4 & \text{for $\omega>\Omega$}
  \end{cases} ~,
\end{equation}
such that the approximate analytic fidelity in Equation (\ref{eq:fidelity_servo}) can be derived. The filter function for amplitude noise is the same for all values of $\phi_c$ such that the fidelity of an arbitrary $\pi$-pulse can be calculated using the filter function
\begin{equation}
    F_\theta^{(\pi)}(\omega)=\left\lvert 1-e^{i\omega\tau_\pi}\right\rvert^2=4\sin^2\left(\frac{\omega\tau_\pi}{2}\right)~,
\end{equation}
which on Taylor expansion can be approximated by 
\begin{equation}
    F^{(\pi)}_{\theta}(\omega)\approx
  \begin{cases}
  \pi^2\omega^2/\Omega^2 & \text{for $\omega<\Omega$}\\
  4 & \text{for $\omega>\Omega$}
  \end{cases}~,
\end{equation}
These Taylor expansions provide the values of $c_b^{(\pi)}$ and $n_\pi$ used in the results.

\begin{acknowledgments}
We would like to thank Virginia Frey for useful discussions on filter functions. This research was supported in part by the Natural Sciences and Engineering Research Council of Canada (NSERC), Grant Nos. RGPIN-2018-05253 and RGPIN-2018-05250, and the Canada First Research Excellence Fund (CFREF), Grant No. CFREF-2015-00011. CS is also supported by a Canada Research Chair.

\end{acknowledgments}

\section*{Data Availability}

The numerical data for generating the plots in this manuscript are available upon reasonable request. Data supplied by third parties is available by consent of the third party.

\bibliographystyle{ieeetr}
\bibliography{lasernoise}

\end{document}